\documentclass{wscpaperproc}
\usepackage{latexsym}
\usepackage{graphicx}
\usepackage{mathptmx}
\usepackage[T1]{fontenc}
\usepackage{amsmath}
\usepackage{amsfonts}
\usepackage{amssymb}
\usepackage{amsbsy}
\usepackage{amsthm}
\usepackage{multicol}
\usepackage{multirow}
\usepackage{subcaption}

\usepackage[pdftex,colorlinks=true,urlcolor=blue,citecolor=black,anchorcolor=black,linkcolor=black]{hyperref}

\newtheoremstyle{wsc}
{3pt}
{3pt}
{}
{}
{\bf}
{}
{.5em}
{}

\theoremstyle{wsc}

\begin{document}

%
%

\pagestyle{fancyplain}

\thispagestyle{plain}
\firstPageHead{}

\chead{\fancyplain{}{\itshape Erazo}}

\rhead{}
\cfoot{}
\renewcommand{\headrulewidth}{0pt} 

\makeatletter
\let\@internalcite\cite
\def\cite{\def\@citeseppen{-1000}%
    \def\@cite##1##2{(##1\if@tempswa , ##2\fi)}%
    \def\citeauthoryear##1##2##3{##1 ##3}\@internalcite}
\def\citeNP{\def\@citeseppen{-1000}%
    \def\@cite##1##2{##1\if@tempswa , ##2\fi}%
    \def\citeauthoryear##1##2##3{##1 ##3}\@internalcite}
\def\citeN{\def\@citeseppen{-1000}%
    \def\@cite##1##2{##1\if@tempswa, ##2)\else{}\fi}%
    \def\citeauthoryear##1##2##3{##1 (##3)}\@citedata}
\def\citeA{\def\@citeseppen{-1000}%
    \def\@cite##1##2{(##1\if@tempswa , ##2\fi)}%
    \def\citeauthoryear##1##2##3{##1}\@internalcite}
\def\citeANP{\def\@citeseppen{-1000}%
    \def\@cite##1##2{##1\if@tempswa , ##2\fi}%
    \def\citeauthoryear##1##2##3{##1}\@internalcite}
\def\shortcite{\def\@citeseppen{-1000}%
    \def\@cite##1##2{(##1\if@tempswa , ##2\fi)}%
    \def\citeauthoryear##1##2##3{##2 ##3}\@internalcite}
\def\shortciteNP{\def\@citeseppen{-1000}%
    \def\@cite##1##2{##1\if@tempswa , ##2\fi}%
    \def\citeauthoryear##1##2##3{##2 ##3}\@internalcite}
\def\shortciteN{\def\@citeseppen{-1000}%
    \def\@cite##1##2{##1\if@tempswa, ##2\else{}\fi}%
    \def\citeauthoryear##1##2##3{##2 (##3)}\@citedata}
\def\shortciteA{\def\@citeseppen{-1000}%
    \def\@cite##1##2{(##1\if@tempswa , ##2\fi)}%
    \def\citeauthoryear##1##2##3{##2}\@internalcite}
\def\shortciteANP{\def\@citeseppen{-1000}%
    \def\@cite##1##2{##1\if@tempswa , ##2\fi}%
    \def\citeauthoryear##1##2##3{##2}\@internalcite}
\def\citeyear{\def\@citeseppen{-1000}%
    \def\@cite##1##2{(##1\if@tempswa , ##2\fi)}%
    \def\citeauthoryear##1##2##3{##3}\@citedata}
\def\citeyearNP{\def\@citeseppen{-1000}%
    \def\@cite##1##2{##1\if@tempswa , ##2\fi}%
    \def\citeauthoryear##1##2##3{##3}\@citedata}
%
%
%
\def\@citedata{%
    \@ifnextchar [{\@tempswatrue\@citedatax}%
                  {\@tempswafalse\@citedatax[]}%
}

\def\@citedatax[#1]#2{%
\if@filesw\immediate\write\@auxout{\string\citation{#2}}\fi%
  \def\@citea{}\@cite{\@for\@citeb:=#2\do%
    {\@citea\def\@citea{, }\@ifundefined
       {b@\@citeb}{{\bf ?}%
       \@warning{Citation `\@citeb' on page \thepage \space undefined}}%
{\csname b@\@citeb\endcsname}}}{#1}}%

%
\def\@citex[#1]#2{%
\if@filesw\immediate\write\@auxout{\string\citation{#2}}\fi%
  \def\@citea{}\@cite{\@for\@citeb:=#2\do%
    {\@citea\def\@citea{; }\@ifundefined
       {b@\@citeb}{{\bf ?}%
       \@warning{Citation `\@citeb' on page \thepage \space undefined}}%
{\csname b@\@citeb\endcsname}}}{#1}}%

%
\def\@biblabel#1{}
\makeatother



\newdimen\bibindent
\bibindent=0.0em
\def\thebibliography#1{\section*{\refname}\list
   {}{\settowidth\labelwidth{[#1]}
   \leftmargin\parindent
   \itemindent -\parindent
   \listparindent \itemindent
   \itemsep 0pt
   \parsep 0pt}
   \def\newblock{}
   \sloppy
   \sfcode`\.=1000\relax}


\setlength{\baselineskip}{12.7pt}

\title{Smart sports predictions via hybrid simulation: NBA case study}

\author{Ignacio Erazo\\[12pt]
	H. Milton Stewart School of Industrial and Systems Engineering \\
	Georgia Institute of Technology\\
	755 Ferst Drive NW\\
	Atlanta, GA 30332-0205, USA\\
}

\maketitle

\section*{Abstract}
Increased data availability has stimulated the interest in studying sports prediction problems via analytical approaches; in particular, with machine learning and simulation. We characterize several models that have been proposed in the literature, all of which suffer from the same drawback: they cannot incorporate rational decision-making and strategies from teams/players effectively. We tackle this issue by proposing hybrid simulation logic that incorporates teams as agents, generalizing the models/methodologies that have been proposed in the past. We perform a case study on the NBA with two goals: i) study the quality of predictions when using only one predictive variable, and ii) study how much historical data should be kept to maximize prediction accuracy. Results indicate that there is an optimal range of data quantity and that studying what data and variables to include is of extreme importance.

\section{Introduction}
Over the last decades, data collection, availability, and usage have greatly increased. Under this context, simulation is used as a powerful tool that allows to optimize the performance of the system being considered. In the past, discrete-event simulation, system dynamics, and agent-based simulation were the main paradigms used, but lately, new paradigms better suited to leverage data by incorporating intelligent frameworks (such as Machine Learning) are gaining steam; in particular hybrid models and metamodels. Prediction is a relevant problem in most areas of study, and while having more data available has facilitated its study through Machine Learning (ML) lenses, solving some problems requires a combination of methods. \shortciteN{Eunji2023} recently authored the paper ``Simulation-based Prediction'', where they consider a hard system with some ``observable data'' that is not necessarily aligned with the full state/data of the system; those difficulties make simulation an essential tool to improve their decisions/predictions. 

Sports prediction is a really hard problem because i) there is a very small amount of ``observable data'' (from games) that does not describe the complete state of the system (health of players, strategies of teams, interactions between players), ii) data is correlated (teams/players compete against each other), and iii) data is time-varying and the ``non-observable'' variables can change significantly over time. To predict the outcomes of sports games, the three following classes of models have been used:

\begin{enumerate}
    \item [1)] Stochastic models: Outcomes of games are computed according to statistical distributions and parameters based on expert knowledge, historical data, and previous studies.
    \item [2)] Play-by-play models: games are modeled as a sequence of plays. By using Markov Chains, the sequence of plays is simulated using the transition probabilities, until the game ends. 
    \item [3)] ML-based models: the independent variables correspond to information known before the game, and the dependent variable is the outcome of the game. Algorithms are trained, then tested over the outcomes of games. 
\end{enumerate}

While the three aforementioned classes of models are concerned with the prediction of the outcome of a game/match, they can be used as a subroutine to simulate complete seasons and tournaments. That can be achieved by considering each game as an event  and simulating games sequentially according to the tournament's rules, essentially following a discrete-event logic. These models can very easily integrate new sources of data to improve predictions, however, they cannot incorporate strategy and game theory over the simulation of games and seasons, unless it is at the expense of an extremely large increase in complexity (i.e., distribution parameters, state space for Markov Decision Processes, number of columns for ML). This is a large drawback, as strategy plays a large role in sports.

To overcome the difficulty of incorporating strategy and decision-making from teams and players, we propose a hybrid simulation model that uses agent-based logic. Our model can also have as a subroutine any of the three aforementioned classes of prediction models for games and generalizes the widely used discrete-event logic. The organization of the paper is as follows: Section \ref{sec:lit_review} presents a literature review on hybrid simulation, the sports prediction problem, some NBA simulation applications, and our contributions. Section \ref{sec:model} introduces our agent-based logic for simulation; whereas Section \ref{sec:case_study_methodology} introduces the specifics and methodology of our case study. Section \ref{sec:case_study} shows and discusses the results of our case study. Finally, Section \ref{sec:conclusions} provides our conclusions and future avenues of research.

\section{Literature Review}\label{sec:lit_review}

\subsection{Hybrid Simulation}

Hybrid simulation has seen an almost exponential increase in usage in the past two decades, and in the state-of-the-art review of \shortciteN{Brailsford2019}, hybrid simulation is referred to as a ``modeling approach combining two of the following: discrete-event simulation, system dynamics, and agent-based simulation''. They also mention that the definition is loose, and is now associated with systems using only one simulation paradigm plus other analytical tools within it, such as machine learning, discrete optimization, and metaheuristics, among others.

Simulation optimization can be seen as a hybrid modeling framework and was described by \shortciteN{Fu2002} as the field that looks for the ``best'' inputs for our decision variables of simulated systems, with ``best'' being roughly equivalent to optimizing the metrics of interest. Nowadays, the field also focuses on optimization problems where solutions lie in a ``policy space'', instead of just a ``decision variable'' space, (see \shortciteN{Fu2014} for a more recent panel talk on simulation optimization). For these problems, the design of the policies is something that needs to be accounted for, just as the optimization of the decision variables (which may vary under different policies). Recent applications of ``policy-space'' problems that can be solved with the help of simulation include logistic problems \shortcite{Santos2016,Lee2018,Erazo2021}, healthcare applications \shortcite{Dorali2022,Erazo2022}, and reinforcement learning problems, among others.

Moreover, hybrid models that incorporate smart decision-making within the simulation have been proposed. These models do not use rule-based approaches to approximate the decisions of agents (e.g., production plan given a realized customer and forecasted demand), or to estimate parameters that influence the outcome of a simulation (e.g., expected time until a component fails). Instead, they use optimization (e.g., optimal production plan) and machine learning algorithms (e.g., predict the time to failure given the current data) as a subroutine to improve the representation of the actual system being simulated. In particular, \shortciteN{Delafuente2018} demonstrate how to enable smart processes within simulation models by integrating Machine Learning within the model's logic, and other applications from digital twins include the use of Neural Networks to improve the representation of real systems \shortcite{Reed2022}.

\subsection{Prediction of Sports Games in a Nutshell}

Prediction of sports games/tournaments has always been an area of active development because results are intricately tied to betting outcomes. In one of the earliest influential works, \shortciteN{Koning2001} predicted the winner of soccer championships using historical data on scoring and adjusting for the quality of opponents. The authors recognize a challenge that still exists: there is not enough data to properly assess the strength level of a team. This is related to what was presented in the introduction, there is ``unobservable'' data, even with all the new sensors and technology. Furthermore, even if there was available a ``strength index'' for each team, there is a need for a probabilistic model that returns, given the strength of both teams in a game, a winning probability for them. It is easy to see how simulation is an attractive solution method for prediction in sports, since different rating models for teams/players and different probabilistic models for outcomes can be tested, and compared to the actual outcomes seen in reality. Recently, \shortciteN{Garnica2022} described a general simulation framework to analyze predictive models for sports with pair confrontations, such as tennis, soccer, football, rugby, basketball, and many others. Their framework includes four sequential steps: i) the competition network, ii) the rating procedure, iii) the forecasting method, and iv) the model validation; they claim that using this methodology may help to provide a better understanding of rating procedures and forecasting techniques for accurate predictions. In the contributions subsection, we present how our hybrid approach enhances this framework.

\subsection{Simulation for NBA Predictions}

The \href{https://www.nba.com/}{National Basketball Association (NBA)} was founded in 1946 and is a major basketball league in the world. Composed of 30 teams with an average value of 2.8 billion USD, it is easy to see its relevance in the sports world, and why many aspects of the game have been researched. In particular, aspects influencing the games' outcomes have been studied, such as the effect of being ``home'' or ``away'' \shortcite{Ribeiro2016}, factors affecting the quality of plays and shots \shortcite{Rolland2020}, the expected value of possessions given tracking data \shortcite{Cervone2016}, among others. The aforementioned aspects affect team performance, which determines games' outcomes.

Concerning simulation towards predicting the games' outcomes, there are two main approaches. In one, the researchers use play-by-play simulation of possessions such as to predict the game outcome. The second approach uses the data of the teams and their players to directly predict the outcome of the game. Concerning the former approach, \shortciteN{Oh2015} considered a data-driven graphical model, \shortciteN{Vracar2016,Sandholtz2018,Sandholtz2020} used Markov-Decision Processes (MDP) to evaluate policies and also model the transitions between plays. For the latter approach, \shortciteN{Song2020} model scores and predict win probabilities based on a Bivariate normal model depending on 5 performance statistics (including ``the four most important factors'', see \shortciteN{Kubatko2007}). Moreover, deep learning methods have been used for the scores' prediction \shortcite{Yanai2022}, but with underwhelming results, very likely due to the small amount of ``observable'' data available.

\subsection{Contributions}

We propose a general hybrid simulation model for seasons/tournaments that enhances the state-of-the-art in simulation for sports predictions by incorporating agent-based aspects. This model allows for smart decision-making of teams/players; leading to a better representation of the system. The model facilitates the integration of steps ii) and iii) from the framework described by \shortciteN{Garnica2022}, but it can also use the two-step method (with a rating step, and a prediction step) presented in that article.

Even with the increased data availability, single metrics remain the norm for communication purposes in sports, as it is easier to compare teams based on a single value. Nonetheless, there exists very limited literature focused on studying the predictive power of these summary statistics for single games. The contributions of our case study are: 
\begin{itemize}
    \item We assess the value of single summary statistics for prediction in the context of NBA games.
    \item For NBA regular season games, we evaluate the following question: ``How much historical/past data should be used to predict the outcome of future games?''. This is a design question that has not been addressed, and results suggest more data is not always better.
\end{itemize}


\section{A Hybrid Model with Agent-Based Components}\label{sec:model}

\subsection{Agents}

\begin{figure}[h!]
    \centering
    \includegraphics[scale=.78]{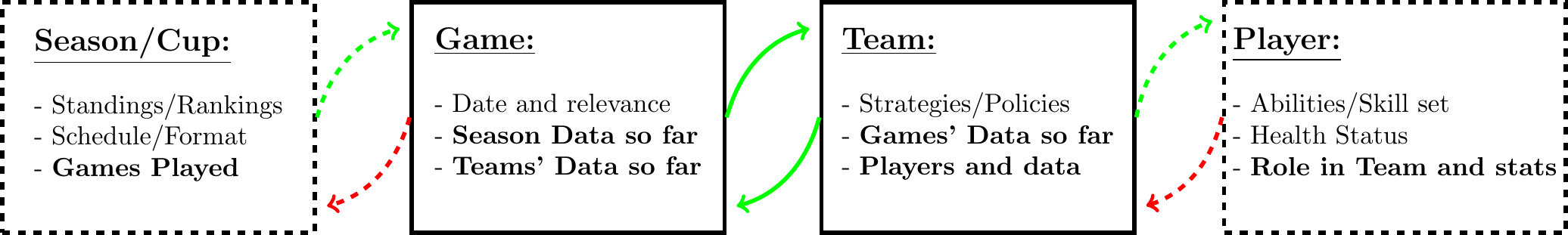}
    \caption{Agents for the hybrid model, and their relationships.}
    \label{Fig:StructureClasses}
\end{figure}

We start by presenting the four classes of agents considered in our simulation model; they are depicted in Figure~\ref{Fig:StructureClasses}. In that figure, the rectangular boxes (Season, Game, Team, Player) represent agents, the green arrows mean a ``one-to-many'' relationship, and the red arrows mean a ``one-to-one'' relationship. Within each agent box, there is some bold text, which corresponds to the information that links the different agents. Some lines and rectangular boxes are dashed; that means those agents and relationships are not imperative for the model to incorporate the agent-based logic.

{}

The only two imperative agents are Games and Teams. Each team is associated with the different games that it must play, and each game carries information about its schedule and relevancy. Teams have strategies and policies, and they react according to the importance of the game being played, and the schedule that surrounds that game. Simulations take into account this behavior, the game being played influences the behavior of teams. The behavior of teams affects the outcome of games, and future games as well (for example, if they decided to not play at full strength in one game, such as to be ready for the next).

Among the optional agents, the Season/Cup agent interacts with the Game agents. By retrieving information from the previous games, the Season agent can modify the relevance of future games. An example of that behavior is when NBA games are moved to national TV (higher importance) or different time slots, depending on the outcome of previous games. Similarly, the outcome of games can affect the rest of the season schedule, for example on championships with elimination games.

Finally, the Player agents interact directly with the Team agents. Teams decide which players get to participate in Games, and the teams' strategies affect the performance of players. Similarly, Players interact with teams by offering their skill sets, but also because they have their own goals. For example, if a team is in a ``contract year'', they benefit from more exposure and hence may push for more minutes in Games, or ask for a trade request. Note that if we are simulating an individual sport, the Team, and Player agents are the same.

\subsection{Simulating a Season/Tournament}

In the previous subsection, we presented the four different agents that may be included in the model, and how they may interact with each other over the course of a simulation. To simulate seasons and tournaments, we still need to simulate each game individually; for which any of the three classes of models presented in the introduction (stochastic models, play-by-play models, or ML-based models) can be used. The interactions between agents will affect the predictions returned by such models by either modifying the parameters for stochastic models, modifying the plays and strategies of teams in the play-by-play models, or modifying the data for which we obtain predictions on the ML-based models. 

After each game, agents have their state variables updated. This includes for the Season agent the current standings, rankings, and schedule. The future Games agents may have their relevance updated. Teams have their previous data updated, positions in the ranking updated, and their strategies and policies may change due to past results. Finally, Player agents also have their statistics updated, their health status updated, and even their future roles on the team may change. The outcomes of each game will depend on all of these state variables, and the end-of-the-season standings will do as well.

The proposed methodology is very broad and general, as it allows to incorporate the previous prediction methods, plus the decision-making of the agents. In particular, if we just set all the teams/players to have no strategy, then simulating a season/tournament just reduces to a discrete-event simulation model, generalizing the previous approaches in the literature. Concerning the general simulation framework to analyze predictive models for sports with pair confrontations presented by \shortciteN{Garnica2022}, we expand the methodology by adding the rational decision-making of agents. Moreover, just like in their framework, our model allows prediction for individual matches with a two-step approach (compute a team rating, then return winning probabilities), but it also allows a one-step approach based on data (predictions based on statistical distributions or ML-based).  

\section{NBA Case study --- Methodology}\label{sec:case_study_methodology}

In our case study, we focus on predicting the binary outcome of NBA \textit{regular season} games. To do so, we downloaded ten seasons of games' box score data (for both teams), starting in the 2011-2012 season and until the season 2021-2022, using the Python module \href{https://github.com/swar/nba_api}{nba\_api}. We then excluded season 2019-2020 from our dataset, because of the inconsistent conditions encountered by teams over that season, such as an extended and unprecedented mid-season break because of Covid-19, then a ``closed-doors'' end-of-the-season competition in a neutral site, and without all the teams and players being present. 

By using many box score statistics, diverse probabilistic models, and/or machine learning algorithms, previous works have achieved an accuracy between 62 and 69\% \shortcite{Manner2016,Song2020} for the binary outcome of NBA games (usually computed over the second half of an NBA season). Previous works predict the outcomes of each game independently, thus when they simulate a season, they are essentially using a Monte Carlo model, instead of a discrete-event model where the simulated outcomes of a game affect the simulated outcome of future games. Having said that, we will also use that logic in our case study, such as to have a consistent benchmark with the previous literature. 

For each season, games will be simulated chronologically. We will predict the outcome of each game using a two-step prediction approach, based on the data available at the beginning of the game. As mentioned earlier, the first step computes a strength ``rating'' for each team, and the second step returns win probabilities based on the parameter of each team in the game. Optimizing the creation of single summary statistics and ``strength'' ratings goes beyond the scope of this work; instead, we will use some widely-available and easy-to-compute single statistics as ratings. The motivation is two-sided: 1) by using simple statistics, we will be able to study their value towards predicting the outcome of games, and 2) this simple setup makes it easy to study a relevant but neglected topic in the sports prediction literature: the quantity of \textit{historical} data (in terms of previous matchups/games) that should be used to make predictions.

\subsection{Ratings for the First Step: Single Summary Statistics}

We present now the simple summary statistics that will be used in our case study. We consider the following two values:

\begin{enumerate}
    \item Win percentage: let $x_1^i$ be 1 if ``Team 1'' won its $i^{th}$ game and 0 otherwise, then the win percentage of ``Team 1'' before the game $i+1$ is: $p_1^i = \dfrac{\sum_{k=1}^i x_1^k}{i}$. For the game $i+1$ we will use the value $\hat{p_1^i} = \dfrac{\pi+\sum_{k=1}^i x_1^k}{1+i}$, where $\pi \in (0,1)$ is just a ``prior'' value that helps to initialize the simulation avoiding trivial win probabilities. $\pi$ can be neutral (0.5) or reflect preseason betting odds.
    \item \href{https://hoopinformatics.medium.com/nba-stats-explained-offensive-defensive-rating-974604afd410}{Net rating:} the net rating of a team before the game $i+1$ corresponds to the sum of points scored minus the sum of the points conceded over games $1, \ldots, i$; scaled by the total number of possessions played in those games. It essentially represents the net efficiency of a team, where a larger number is better. We initialize simulations with a value of 0 for each team.
\end{enumerate}

For both input values, we refer to them as ``home-adjusted'' if the value used for the team playing at ``home'' is computed only considering its respective ``home'' games; while the overall value (considering all games) is used for visiting teams.

\subsection{Functions for the Second Step} \label{subsec:funcions}

In the previous subsection, we presented the two summary statistics that will be used as the output of the first step of the prediction procedure. Now, we go over the functions that will be used in the second step. For ease of exposition, we will just refer to teams, but they may represent players in individual sports. We assume that ``Team 1'' plays against ``Team 2''.

\subsubsection{Real-Valued Rating}
Let's assume that the output of the first step is a real-valued number and that a larger value means better. The simplest procedure just returns deterministically the team/player with the highest value as the winner (if there is one), or a tie if both teams have the same value (for sports with no ties, it returns instead a 50\% win chance of winning for both teams). 
    
While simple, this method has a desirable property: the accuracy should increase if the rating input is a better representation of the team/match conditions. For our case study, this function can be used with both win percentage, and net rating as inputs. 

\subsubsection{Probability-Valued Rating}

Now, let's assume the input is a ``probability-valued'' index between 0 and 1; for example, the probability of winning versus an ``average team''. We define the \textit{Bernoulli Race} between ``Team 1'' and ``Team 2'' with parameters $p_1, \ p_2$ (respectively) as the distribution denoted $BR(p_1, \ p_2)$ that returns:
    \begin{itemize}
        \item For sports allowing ties: ``Team 1'' wins with probability $p_1(1-p_2)$, ``Team 2'' wins with probability $(1-p_1)p_2$ and there is a tie with probability $(1-p_1)(1-p_2) + p_1p_2$. Intuitively, each team performs a Bernoulli trial according to their parameter, and if the obtained values are equal then it is a tie, otherwise, the team that obtains the positive outcome wins.
        \item For sports without ties: ``Team 1'' wins with probability $\frac{p_1(1-p_2)}{p_1(1-p_2) + (1-p_1)p_2}$, ``Team 2'' wins with probability $\frac{(1-p_1)p_2}{p_1(1-p_2) + (1-p_1)p_2}$. Intuitively, ``Team 1'' and ``Team 2'' draw Bernoulli samples until one of them gets a value of one (wins) and the other does not (losses).
    \end{itemize}

This distribution is particularly interesting for sports with no ties because of the following property: if we let $p$ be the probability of beating an ``average'' team, then an average team would have a value of $p=0.5$. If ``Team 1'' plays versus an average ''Team 2'' with $p_2=0.5$, then the probability of ``Team 1'' winning is exactly $\frac{0.5p_1}{0.5p_1 + 0.5 - 0.5p_1} = p_1$; the interpretation of $p$ values remains consistent. Also, if $p_1=p_2$, both teams have a 0.5 probability of winning their match; again ensuring consistency. Furthermore, the interpretation of the $0.5$ value makes sense in sports where each team plays the same amount of games because, under those circumstances, the average win ratio (the average of the ratios of each participant) is exactly equal to 0.5. For our case study, this function can be used with the win percentage as input.

\subsection{Simulation Models}

In our case study, we consider the following six methods for the prediction of games:

\begin{multicols}{2}
\begin{enumerate}
    \item [(i)] \textit{Bernoulli Race}, based on win percentage.
    \item [(ii)] Home-adjusted \textit{Bernoulli Race}, based on win percentage.
    \item [(iii)] Largest value, based on win percentage.
    \item [(iv)] Home-adjusted largest value, based on win percentage.
    \item [(v)] Largest value, based on net rating.
    \item [(vi)] Home-adjusted largest value, based on net rating.
\end{enumerate}
\end{multicols}

We will use two different simulation models: a ``basic'' version, and an ``expanded'' version. In the ``basic'' model, the single summary statistic (win percentage or net rating, either overall or ``home-adjusted'', depending on the method (i)--(vi) used) is computed before each game for both the home team and the away team, and the game outcome depends solely on the summary statistics. On the other hand, the ``expanded'' model also integrates rational decision-making by the teams, which affects the outcome of games.

In the ``expanded'' model, the teams are always aware of the league standings, and their playoff status (cannot classify, fighting for a playoff berth, or classified). Furthermore, teams are also aware of the draft picks that they own for the next NBA Draft (information obtained from \href{https://www.prosportstransactions.com/basketball/DraftTrades/Years/index.htm}{Pro Sports Transactions}). Teams that do not classify for the playoffs participate in the \href{https://en.wikipedia.org/wiki/NBA_draft_lottery}{NBA Draft Lottery}, and they have greater odds of being awarded the best selection if their record is worse. Because of this, teams that own their draft pick in the upcoming NBA Draft have an incentive to lose after being eliminated from playoff contention. On the other hand, teams that are already classified for the playoffs have an incentive to rest their players in the last few games, so they avoid injury and start the playoffs rested. 

We added those two incentives into the simulation using a simple rule-based approach:
\begin{itemize}
    \item If a team owns its first-round pick in the upcoming NBA Draft and is already eliminated from playoff contention, then on methods (i)--(iv) its winning percentage (either overall or home-adjusted) is reduced by half. For methods (v) and (vi); its net rating is reduced by 5. 
    \item If a team has to play at most three extra games in the regular season, and is already classified for the playoffs, then its winning percentage for methods (i)--(iv) (either overall or home-adjusted) is reduced by half. For methods (v) and (vi); its net rating is reduced by 5.
    \item The \href{https://en.wikipedia.org/wiki/NBA_play-in_tournament}{Play-in Tournament} was added in the season 2019-2020 as a method to reduce the ``losing'' incentives by making the playoff more exciting and unpredictable. During the 2012-2019 period, teams ranked 1-8 in each conference would classify for the playoffs. After the inclusion of the play-in games, teams 1-6 in each conference would classify directly, and teams 7-10 would participate in the Play-in Tournament to decide the last two seeds. For seasons 2012-2013 up to 2018-2019, we consider a team as ``classified'' if it that cannot end lower than eighth in the standings, and a team is ``eliminated'' if it cannot end higher than ninth. For seasons 2020-2021 and 2021-2022, a team is considered as ``classified'' if it cannot end the season lower than sixth in the standings, and ``eliminated'' if it cannot end higher than eleventh.
\end{itemize}

The choice of halving the win percentage, and the choice of reducing the net rating by five, were arbitrary. As the incentives represent logical decisions from a team perspective, we speculated that any ``reasonable'' choice of incentive parameters was going to help to improve the prediction accuracy. We decided not to perform any tuning to the incentive parameters, such as to avoid the need of excluding a few seasons to have both a training and a testing set.

\section{Case study --- Results}\label{sec:case_study}

We first compare the accuracy obtained by each of the six methods (i)--(vi) for our ``basic'' simulation model. We simulated each of the nine seasons one thousand times, and the aggregated results are presented in Table~\ref{Tab:AccuracyResultsBasic}. Comparisons 1, 2, and 3 help to visualize the difference between the regular parameters, and the ``home-adjusted'' parameters. It is easy to see that for the \textit{Bernoulli Race} probabilistic model (Comparison 1), using the ``home-adjusted'' parameter increases the prediction accuracy on the nine simulated seasons, for both the complete season and also for the second half part of the season. Furthermore, the overall prediction accuracy increases by at least 1\% for the complete season, and for the second half of it. Concerning the largest value model (Comparisons 2 and 3), the use of the ``home-adjusted'' parameter (win percent for Comparison 2, and net rating for Comparison 3) does not appear to have major effects; with overall differences below 0.2\% across the nine seasons.

Concerning the performance of the different methods, Table \ref{Tab:AccuracyResultsBasic} also shows that all the methods have a larger accuracy when predicting only the outcome of games in the second half of the schedule. This is very likely due to the fact that predicting the winner of the first few matches of the season is particularly hard as there is a lack of data, and teams are still getting into their ``in-season'' form. The (almost completely) deterministic largest value method performs significantly better, reaching prediction accuracies of at least 66\% for methods (iii)--(vi) during the second half of the seasons; and the use of the net rating parameter seems to have an edge, with accuracies of at least 66.5\%. These results suggest that our ``basic'' model can achieve a prediction accuracy similar to the standard in previous works, but just using a simple deterministic model with single easy-to-compute values. It follows that, even in the presence of just a few variables, decently good predictions can be obtained if meaningful information is provided.

\begin{table}[h!]
\caption{Average accuracy of games across the nine NBA regular seasons simulated (each 1000 runs), depending on the method used, for the ``basic'' simulation model.}
\label{Tab:AccuracyResultsBasic}
\centering
\begin{tabular}{|c|c|c|c|||c|c|||c|c|||}
\cline{3-8}
\cline{3-8}
\multicolumn{2}{c}{} & \multicolumn{6}{|||c|||}{Method used}\\
\cline{3-8}
\cline{3-8}
\multicolumn{2}{c}{} & \multicolumn{2}{|||c|||}{Comparison 1} & \multicolumn{2}{|c|||}{Comparison 2} & \multicolumn{2}{|c|||}{Comparison 3}\\
\cline{1-8}
Metric of interest & Interval of season & \multicolumn{1}{|||c|}{(i)} & (ii) & (iii) & (iv) & (v) & (vi)\\
\hline
\multirow{2}{*}{Prediction accuracy} & Complete season & \multicolumn{1}{|||c|}{56.9\%} & 57.9\% & 64.1\% & 64.3\% & 63.8\% & 63.7\%\\
\cline{2-8}
 & 2nd half & \multicolumn{1}{|||c|}{57.3\%} & 58.6\% & 66.0\% & 66.0\% & 66.7\% & 66.5\%\\
 \hline
Higher mean accuracy & Complete season & \multicolumn{1}{|||c|}{0} & 9 & 3 & 6 & 6 & 3\\
\cline{2-8}
for a specific season & 2nd half & \multicolumn{1}{|||c|}{0} & 9 & 4 & 5 & 6 & 3\\
\hline
\end{tabular}
\end{table}

We now compare the performance of the ``basic'' and ``extended'' models. Table~\ref{Tab:AccuracyResultsExtended} shows the results after simulating 1000 times each of the nine seasons, for each of the methods (i)--(vi), and using the ``extended'' model. By comparing against the values from the ``basic'' model, shown in Table~\ref{Tab:AccuracyResultsBasic}, it is easy to see that each of the methods improves its respective overall prediction accuracy when switching to the ``extended'' model. The improvement is larger in the second half of the season, and that follows from the fact that the losing incentives are stronger towards the end of the season when standings are almost decided. This consistent improvement suggests the potential of adding rational decision-making from agents to predict the outcome of sports games. Moreover, when looking at each of the Comparisons 1, 2, and 3; the same behavior occurs, the ``home-adjusted'' parameters perform better when using the \textit{Bernoulli Race} method, and they perform similarly to their overall counterpart when used for methods (iii)--(vi).

\begin{table}[h!]
\caption{Average accuracy of games across the nine NBA regular seasons simulated (each 1000 runs), depending on the method used, for the ``extended'' simulation model.}
\label{Tab:AccuracyResultsExtended}
\centering
\begin{tabular}{|c|c|c|c|||c|c|||c|c|||}
\cline{3-8}
\cline{3-8}
\multicolumn{2}{c}{} & \multicolumn{6}{|||c|||}{Method used}\\
\cline{3-8}
\cline{3-8}
\multicolumn{2}{c}{} & \multicolumn{2}{|||c|||}{Comparison 1} & \multicolumn{2}{|c|||}{Comparison 2} & \multicolumn{2}{|c|||}{Comparison 3}\\
\cline{1-8}
Metric of interest & Interval of season & \multicolumn{1}{|||c|}{(i)} & (ii) & (iii) & (iv) & (v) & (vi)\\
\hline
\multirow{2}{*}{Prediction accuracy} & Complete season & \multicolumn{1}{|||c|}{57.3\%} & 58.2\% & 64.1\% & 64.5\% & 64.0\% & 64.0\%\\
\cline{2-8}
 & 2nd half & \multicolumn{1}{|||c|}{58.0\%} & 59.2\% & 66.2\% & 66.4\% & 67.1\% & 67.0\%\\
 \hline
Higher mean accuracy & Complete season & \multicolumn{1}{|||c|}{0} & 9 & 3 & 6 & 5 & 4\\
\cline{2-8}
for a specific season & 2nd half & \multicolumn{1}{|||c|}{0} & 9 & 3 & 6 & 5 & 4\\
\hline
\end{tabular}
\end{table}

\begin{figure}[h!]
    \centering
    \includegraphics[scale=1.0]{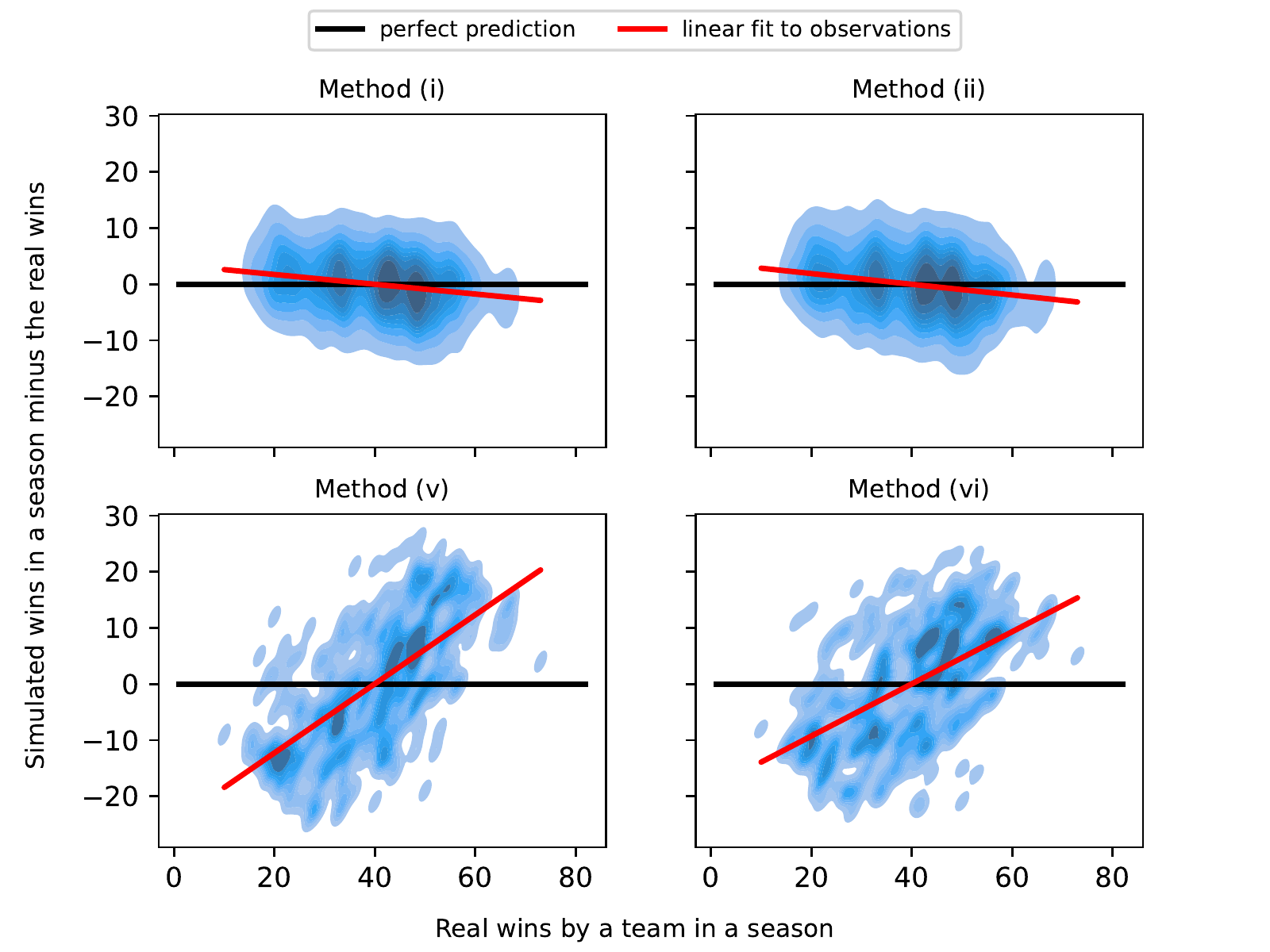}
    \caption{Distribution of Simulated wins - Real wins for methods (i), (ii), (v), and (vi), over the nine NBA regular seasons considered in the case study, and using the ``extended'' simulation model.}
    \label{Fig:OfflineDist}
\end{figure}

Our previous analysis highlights the quality of our predictions for both the ``basic'' and ``extended'' simulation models. It is easy to see that the accuracies obtained match the standards found in the literature when using the largest value method, but the same is not true when using the \textit{Bernoulli Race} probabilistic model. The increase in overall prediction accuracy comes from having deterministic outcomes for most of the games (unless the parameters for each local and visit team are equal), and that leads to the trade-off of having more unbalanced outcomes over the simulation records or having worse overall accuracy of prediction. Figure~\ref{Fig:OfflineDist} shows the distribution of (simulated wins - real wins) for all teams, over the course of all the replications and for the nine NBA seasons that were simulated, using the ``extended model''. The x-axis shows the number of real wins obtained by the team in the respective season, whereas the y-axis shows the difference between the number of simulated wins and the number of real wins. Stronger blue colors represent a larger concentration of observations, the black horizontal segment represents what perfect predictions would look like, and the red segment is the linear fit to all the observations (trend line).

{}

It is easy to see that the \textit{Bernoulli Race} methods lead to more balanced outcomes compared to the larger value methods. Methods (i) and (ii) have similar performance, with a slight inclination to increase by a bit the number of wins of bad teams and decrease by a bit the number of wins of good teams. On the other hand, the largest value methods (v) and (vi) are extremely biased towards good teams, as their number of wins is consistently overestimated. This behavior comes from the fact teams with large net ratings are guaranteed to win under the deterministic model, and the opposite happens for teams with low net ratings. We also note that method (vi) performs better than method (v) with respect to the distribution inequality, and that hints at an advantage of using the ``home-adjusted'' parameter instead of the overall counterpart.

\subsection{How Much Historical Data Should be Kept?}

\begin{figure}[h!]
    \centering
    \includegraphics[scale=1]{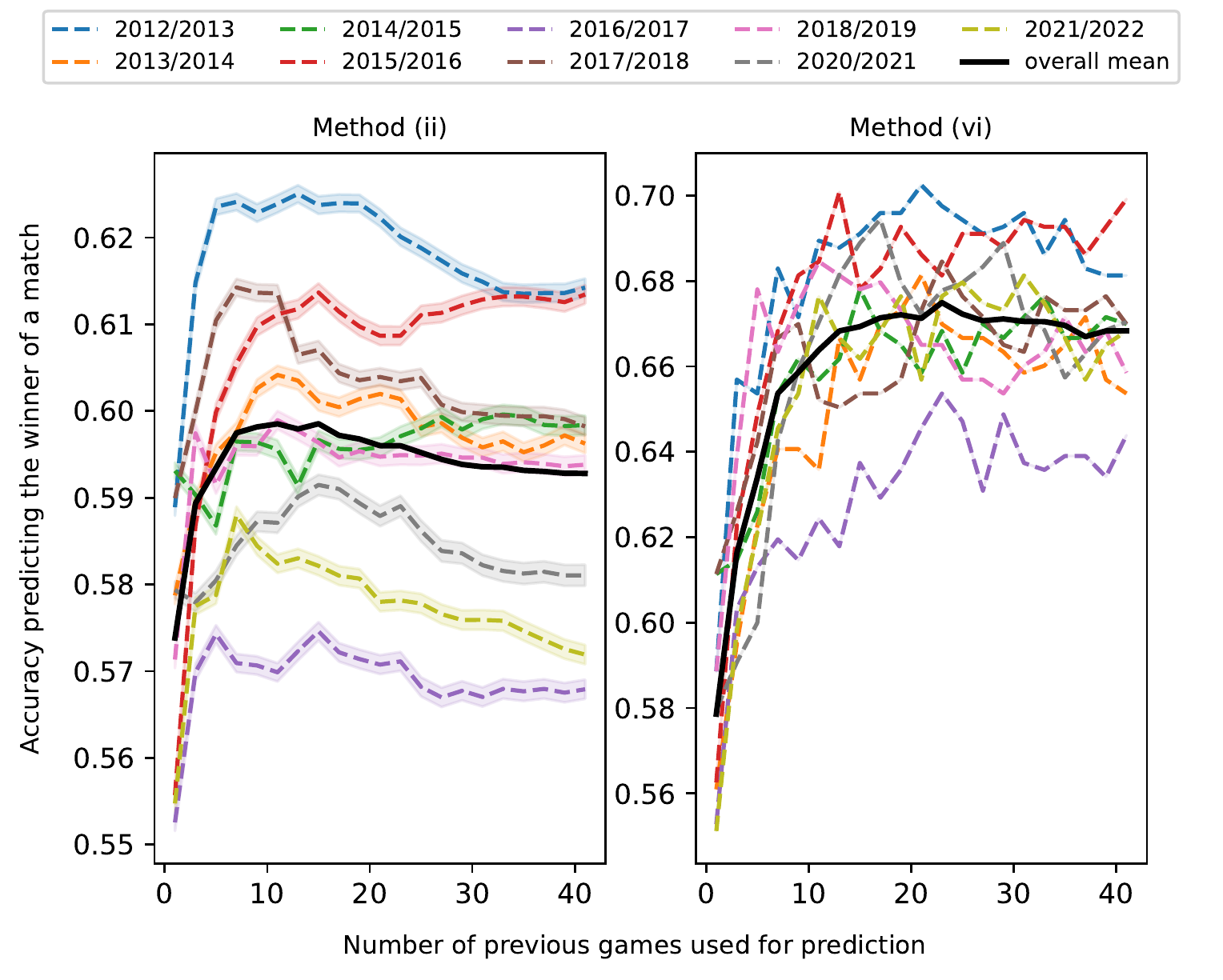}
    \caption{Impact of the number of games used to compute the respective single summary statistics for methods (ii) and (vi), when simulating with the ``extended'' model. The prediction accuracy is reported for nine NBA regular seasons, simulated 1000 times each.}
    \label{Fig:Lags2nd}
\end{figure}

We now study the relationship between the amount of historical data used for prediction and the overall accuracy of the predictions. We simulated each of the nine seasons in our dataset one thousand times, using methods (ii) and (vi) with the ``extended'' simulation model.
Figure~\ref{Fig:Lags2nd} shows the results. The x-axis represents the number of games kept in the data for prediction, and each of the colored curves represents the prediction accuracy of the respective method given the specific number of games kept for prediction purposes. The colored area around the curves represents the 95\% confidence interval on the prediction accuracy; note that the curves from method (vi) do not have colored areas, this follows from the fact that the predictions given by that method are almost completely deterministic. The black curve corresponds to the overall prediction accuracy considering all the games simulated. 

From Figure~\ref{Fig:Lags2nd}, we notice that both methods exhibit similar behavior. Initially, prediction accuracy improves when adding more games, then gets to an interval where it hits the maximum accuracy, afterwards, it starts dropping slowly when adding more games. In particular, method (ii) achieves its maximum accuracy when using the last 8--15 games of data for prediction. Method (vi) achieves its maximum performance when using between 18 and 25 games of data. This behavior also occurs when evaluating the other methods, and when considering the ``basic'' simulation model. These results suggest that there may be an optimal range of historical data that should be kept to predict the outcome of games and that such range depends on the methodology used for prediction, and on the particular sport/competition that is simulated. We encourage research on this topic, because it may lead to improvements in the performance of models that predict the outcome of sports games. 

{}

\section{Conclusions}\label{sec:conclusions}

In this article, we described the sports prediction problem and emphasized the complexity of getting accurate prediction results. In particular, we describe the current methods used to predict the winner of sports games/matches, and how those methods can be used as a subroutine when trying to simulate a season or tournament, which can be loosely seen as a discrete-event simulation. We then proposed a hybrid simulation model that adds agent-based logic to the sports prediction problem, allowing the introduction of strategy and rational thinking from teams and players. 

We performed a case study on the prediction of NBA regular season games. In order to do that, we used two very simple and widely-available summary statistics and considered two simulation models, a ``basic'' model that does not incorporate rational thinking of teams and an ``extended'' model that does. We had two main objectives: evaluate the power of single summary statistics for prediction, and study the relationship between the number of previous games kept for prediction, and the overall accuracy obtained.

Results indicate that even under single-valued inputs and simple functions, similar accuracies compared to data-intensive algorithms can be achieved; however, it comes at the expense of a large positive (negative) bias towards good (bad) performing teams. This suggests that more effort needs to be done on identifying key measures influencing the success probabilities for teams, rather than using as much data as possible. Moreover, the ``extended'' model outperformed slightly the ``basic'' model, which hints at the fact that incorporating more rational thinking and strategy from teams has the potential to improve the quality of models predicting the outcome of sports games. Furthermore, our results indicate that including more historical data for prediction is not always better and that there may be an optimal amount of data to use. This is an area that should be explored more in the future. 

Finally, with respect to future research, it would be interesting to study the performance of models that leverage the agent-based components at their full strength, including the Player agents and their behavior. In addition, including more data to perform the predictions would be an interesting research direction, such as to evaluate how much the prediction accuracy can increase. Furthermore, evaluating the performance of the model when predicting the outcome of other sports is also something worth pursuing.

\section*{ACKNOWLEDGMENTS}
The author thanks the five anonymous reviewers and the committee members for their thoughtful comments and suggestions. He also thanks Prof. David Goldsman for his encouragement and support in this work. 


\footnotesize

\bibliographystyle{wsc}

\bibliography{demobib}

\section*{AUTHOR BIOGRAPHIES}

\noindent {\bf IGNACIO ERAZO} is a Ph.D. Candidate in Operations Research in the H. Milton Stewart School of Industrial and Systems Engineering at the Georgia Institute of Technology. He is focused on efficient data-driven decision-making via integer optimization and large-scale simulation optimization. His email address is \email{ierazo@gatech.edu} and his website is \url{https://ierazo.github.io}.

\end{document}